\documentclass[conference,compsoc]{IEEEtran}

\usepackage{url}
\usepackage[hidelinks]{hyperref}
\usepackage{xcolor}
\usepackage{balance}

\definecolor{darkcyan}{rgb}{0.0, 0.55, 0.55}
\definecolor{darkpastelred}{rgb}{0.76, 0.23, 0.13}

\newcommand{\todo}[1]{{\color{red}--TODO}\xspace}

\begin{document}

\title{Auditing the Impact of Cross-Site Web Tracking on YouTube Political and Misinformation Recommendations}

\author{
\IEEEauthorblockN{Salim Chouaki, Savaiz Nazir, Sandra Siby}
\IEEEauthorblockA{New York University Abu Dhabi, Abu Dhabi, UAE\\
Emails: \{salim.chouaki, an3909, sandra.siby\}@nyu.edu}
}

\maketitle

\begin{abstract}

YouTube has today become the primary news source for many users, which raises concerns about the role its recommendation algorithm can play in the spread of misinformation and political polarization. Prior work in this area has mainly analyzed how recommendations evolve based on users' watch history within the platform. Nevertheless, recommendations can also depend on off-platform browsing activity that Google collects via trackers on news websites, a factor that has not been considered so far. To fill this gap, we propose a sock-puppet-based experimental framework that automatically interacts with news media articles and then collects YouTube recommendations to measure how cross-site tracking affects the political and misinformation content users see. Moreover, by running our audits in both tracking-permissive and tracking-restrictive browser environments, we assess whether common privacy-focused browsers can protect users from tracking-driven political and misinformation bubbles on YouTube.

\end{abstract}

\section{Introduction }

During the past decade, there has been a shift in the way people encounter news-related and political information on the web. Rather than deliberately selecting and visiting a small number of news outlets, many users now consume news as a by-product of their activity on online platforms. YouTube in particular is an important source of information: a Pew Research Center study reports that almost half of U.S. adults get news from YouTube~\cite{pew_social_media_news_2025}, and YouTube's Chief Product Officer states that 70\% of videos watched on the platform come via its recommendation algorithm~\cite{cnet_youtube_ces2018}. This raises questions about the implications of the recommendation algorithm. Although it is designed to personalize recommendations to each user~\cite{milano2020recommender}, it can also contribute to the spread of misinformation and to the formation of filter bubbles and echo chambers~\cite{pariser2011filter, bozdag2015breaking, roose2019making}.

As a result, a large body of work has studied whether YouTube's recommendation system amplifies political polarization or misinformation~\cite{chen2023subscriptions, hosseinmardi2021examining, ledwich2019algorithmic, lewis2018alternative}. Previous work shows that YouTube recommendations can pull users away from extreme content and that recommendations skew left even when there is no watch history~\cite{10.1093/pnasnexus/pgad264}, while other work shows that YouTube's algorithm recommends ideologically aligned content to partisan users~\cite{doi:10.1073/pnas.2213020120}. However, most previous work takes a closed-platform view. These studies design experiments where synthetic users watch a set of left- or right-leaning videos on YouTube and then measure how recommendations shift as the watch history evolves.

In reality, YouTube’s recommendation system can rely on more signals than just watch history. For example, Google’s Privacy Policy explains that it collects information about users’ activity on third-party sites and apps that use Google services (e.g., Google Analytics or embedded YouTube videos) and that this information can be used to provide personalized services such as content and recommendations across Google products~\cite{google_privacy,google_partner_sites}. Therefore, off-platform browsing on news websites that include Google services may influence what content users see on YouTube. However, existing studies on political bias and misinformation in YouTube recommendations do not consider these off-platform data that Google has about users. This is particularly important given that Google Analytics alone is used by about 45\% of all websites~\cite{w3techs_ga}, that Google trackers are installed on 72\% of sites~\cite{kieserman2025tracker}, and that news websites tend to include more trackers than other types of sites~\cite{10.1145/2976749.2978313}.

In this work, we aim to address this gap by measuring the effect of off-platform browsing activity on the political and misinformation content of YouTube recommendations. Specifically, we ask three questions. First (\textbf{RQ1}), do off-platform visits to misinformation articles increase the presence of videos related to known false claims in YouTube recommendations? Second (\textbf{RQ2}), does off-platform browsing of left-, right-, or extreme news articles shift the political balance of recommended videos? Third (\textbf{RQ3}), do privacy-enhancing browser configurations that block trackers and third-party cookies reduce or eliminate these tracking-induced shifts in political and misinformation exposure?

Answering these questions will shed light on whether cross-site tracking by online platforms affects the news content users are exposed to. Rather than considering algorithmic ``filter bubbles'' as platform-specific phenomena, this work adopts a web-wide perspective in which off-platform data flows across services contribute to content curation. Moreover, by comparing privacy-enhanced and non-enhanced browsing environments, the study aims to evaluate whether privacy protections mitigate the curation of politically polarized and misinformation-related content.



\section{Experimental methodology}

In order to measure whether off-platform browsing activity has an impact on YouTube's recommendations, we conduct a systematic audit of the platform using automated browser instances or \textit{sock puppets}.
We use sock puppets to mimic users who visit news articles and then get YouTube homepage recommendations.
Our audit has three phases: (1) a setting phase, where we create sock puppets with a fresh browser profile, (2) an exposure phase, where sock puppets visit news articles associated with different political or misinformation categories, and (3) a measurement phase, where we collect their recommendations on YouTube.

\vspace{5pt} \noindent
\textbf{Setting phase.}
At the start of an experiment, each sock puppet uses a fresh browser profile with cleared browser storage. We observed that YouTube does not present recommendations on the homepage until at least one video has been watched. Therefore, the puppet first visits YouTube and watches a sports-related video. 
During this step, YouTube stores identifiers in the browser storage, which may be used to re-identify the user in subsequent phases. We also record homepage recommendations and use them as baseline recommendations.

\vspace{5pt} \noindent
\textbf{Exposure phase.}
We consider six groups of sock puppets and expose each group to a different type of news article to simulate users with varying political or misinformation exposure. We consider four groups to represent different political ideologies—extreme-left, left, right, and extreme-right; one group to represent high exposure to misinformation content; and one control group that does not visit any news article URLs.
To select news article URLs for political-ideology exposure, we use Media Bias/Fact Check (MBFC)~\cite{mediabiasfactcheck}, which labels news outlets by political bias. For each of the four political ideologies we consider, we select 50 outlets that MBFC classifies as having that ideology. From each outlet, we collect 20 news article URLs, resulting in a pool of 1,000 articles per ideology. Moreover, to construct the misinformation articles set, we use PolitiFact~\cite{politifact}, a fact-checking website. We collect all news articles published between 2020 and 2025 that PolitiFact tags as misinformation, yielding a pool of 52,000 articles.

For each sock puppet, we randomly sample 20 articles from the pool corresponding to their group and visit them sequentially. On each news article page, we attempt to accept consent banners to ensure that tracking is allowed~\cite{duckduckgo_autoconsent}, scroll randomly through the article, and remain on the page for up to one minute.

\vspace{5pt}\noindent
\textbf{Measurement phase.}
After the exposure phase, sock puppets return to the YouTube homepage and collect the list of recommended videos shown there. We repeat the exposure and measurement steps once per day over five consecutive days. We then plan to measure misinformation exposure and the political bias of recommended videos for each sock puppet, before and after exposure, and over days,  to examine how recommendation changes emerge and persist over time. We initially focus on misinformation labeling and leave the labeling of videos’ political biases for future work.

To detect misinformation-related videos, we follow an approach inspired by prior work on matching content to fact-checked claims~\cite{10.1093/pnasnexus/pgad264}. We compare each video’s title and transcript to a corpus of fact-checked claims labeled as misinformation by PolitiFact between January 2020 and October 2025. For each video and each fact-checked claim, we compute text embeddings using MPNet~\cite{song2020mpnet}, a pretrained transformer model. We then compute the cosine similarity between the embedding of each video and the embeddings of all misinformation claims, and compare the resulting similarity distributions across experimental settings.

\vspace{5pt} \noindent
\textbf{Experimental environments.}
We conduct each experiment within two different browser environments. The first is a tracking-permissive environment, where there is no protection against tracking. We use this environment to measure the effect of tracking data on content recommendations (RQ1 and RQ2). The second is a tracking-restrictive environment, where users are protected from third-party tracking, which we use to test RQ3. In both environments, sock puppets are not logged into any Google account.

The tracking-permissive environment is the Google Chrome browser with default settings. In this configuration, third-party cookies are allowed and third-party tracking is not blocked. Therefore, Google should be able to collect users' browsing activity from all third-party websites that include a Google service.
The tracking-restrictive environment is the Brave browser, with privacy protections enabled (i.e., Shields up). In this configuration, third-party cookies are blocked and Brave blocks third-party tracking requests by default. Hence, Google should not be able to observe users' browsing activity on third-party websites.

\section{Preliminary Results}

We measured the effect of tracking on misinformation exposure. We deployed 100 sock puppets and collected a total of 23,854 recommended videos before training and 27,162 after training. In a non-private environment, recommendations become more similar to known misinformation claims after training (+0.0134 cosine similarity to PolitiFact false-claims), whereas no such increase appears in a private browsing environment. 
This suggests that off-platform interactions with news might influence the news content recommended by YouTube, when tracking is allowed.

\section{Future work}

We aim to extend this work in several avenues. 
First, beyond misinformation exposure, we plan to analyze how visits to different news articles affect the political bias of news recommendations on YouTube. 
Second, we plan to study the temporal dynamics of personalization by varying the time between visiting news articles and collecting recommendations, to measure how quickly recommendations change and how long these effects persist. Finally, we will compare logged-in and logged-out settings by auditing recommendations for sock puppets that are signed into Google accounts versus anonymous ones.

\section*{Acknowledgment}
This work was supported by Tamkeen under the NYU Abu Dhabi Faculty Research Funds.

\balance
\bibliographystyle{plain}
\bibliography{bibliography/references, bibliography/links}

\end{document}